\documentclass[aps,twocolumn,raggedbottom,prl,showpacs,epsfig]{revtex4}
\usepackage{epsfig}
\begin{document}
\title{Multi-band quantum ratchets}
\author{M. Grifoni$^{1}$, M.~S. Ferreira$^{1,2}$, J. Peguiron$^{1}$, and J.~B. Majer$^{1}$}
\date{\today}
\affiliation{$^{1}$  Department of Nanoscience, Delft University of Technology,
 Lorentzweg 1, 2628 CJ Delft, The Netherlands \\
$^{2}$ Department of Physics, University of Dublin, Trinity College, Dublin 2,
Ireland\\   }
\begin{abstract}
We investigate  directed motion in non-adiabatically rocked ratchet systems sustaining few bands below the barrier.
Upon restricting the dynamics to the lowest $M$ bands, the total system-plus-bath Hamiltonian is mapped onto a discrete tight-binding model containing all the information both on the intra- and inter-well tunneling motion.
A closed form for the current in the incoherent tunneling regime is obtained.
In effective single-band ratchets, {\em  no}  current rectification occurs.
We apply our  theory  to describe  rectification effects in
  vortex quantum ratchets devices. Current reversals upon variation of the ac-field amplitude or frequency are predicted. 
 \end{abstract}
\maketitle
%
 A ratchet, i.e. a periodic structure with broken spatial symmetry, yields the possibility to obtain a directed current in the presence of noise and 
 unbiased nonequilibrium forces \cite{Reimann01}. As such,
 ratchets belong to the class of so termed Brownian motors \cite{Special02}, 
 being devices
 operating far from equilibrium, and which combine noise and asymmetry to
 generate particle's transport.
  A huge amount of experimental and theoretical work exists on ratchets ruled
by classical thermal fluctuations \cite{Reimann01,Special02}, also due to their possible application in biological systems \cite{Maddox93,Astumian97}. In contrast, little is known
about the ratchet effect in the  quantum realm. This is partly  due to the challenge in the experimental realization of suitable ratchet potentials. 
 Only recently   rectification of
quantum  fluctuations
in  triangularly shaped  semiconductor heterostructures \cite{Linke99}, and   in quasi  one-dimensional  Josephson junction arrays \cite{Majer01} has been 
 observed.
Theoretically, this originates from the complexity in the investigation  of the  quantum ratchet dynamics.    By now   the quantum ratchet effect has been
 tackled  only for  adiabatically rocked ratchets in a mostly numerical work \cite{Reimann97},
 for peculiar single band models \cite{ratch2}, 
  in the presence of an external force of on-off type   \cite{ratch1}, and for a weakly corrugated potential \cite{Scheidl02}.  Typical quantum features, as e.g. a current inversion upon temperature decrease   were  predicted \cite{Reimann97} and demonstrated \cite{Linke99}. So far, the band structure of the potential was not considered.

In this letter for the first time
 a microscopic theory for the interplay among tunneling, vibrational relaxation
  and
  non-adiabatic driving in  ratchet potentials sustaining few bands below the barrier, cf. Fig. 1, is presented.  For these potentials the semiclassical requirement \cite{Reimann97} of having many bands below the barrier is not met. 
 Our  treatment, mostly analytical, is based on the real time path-integral formalism for open quantum systems \cite{Weiss99}.
Noticeably, in the temperature and driving regime in which the dynamics is effectively restricted to
 the lowest band of the periodic potential, {\em no} current rectification occurs.  In fact,
 a reduction to the  lowest band  of the  ratchet potential  {\em only} retains information about the periodicity
 of the original Hamiltonian, but not about its reflection properties.
 To  take into account the vibrational motion within the well, and hence the asymmetry of the ratchet potential leading to the ratchet effect, {\em at least two bands} should contribute.
\begin{figure}
\epsfig{figure=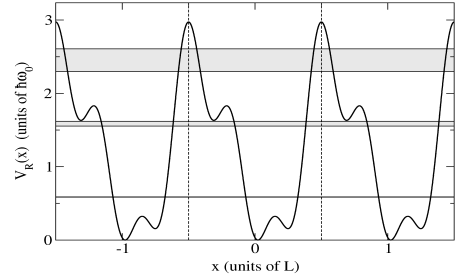,width=45mm,height=75mm,angle=270}
\caption{Ratchet potential   with three bands below the barrier (shaded regions). The potential height is in units of the
 distance $\hbar\omega_0$ between the centers of the second and first band, and the length  is  in
units of the period $L$.   }
\end{figure}
Starting from the system-plus-reservoir model of   a driven ratchet system bi-linearly coupled to a harmonic oscillator bath,   we proceed in three steps: i) We consider the regime of temperature and driving parameters which allows the truncation of the full dynamics to the Hilbert space spanned by the $M$ lowest bands of the  periodic ratchet potential. ii) A rotation to the so-called discrete variable representation (DVR) \cite{DVR}, being the eigenbasis of the discrete position operator coupling to the bath, is performed. In this basis, the isolated ratchet problem is described by a tight binding model with a periodicity of $M$ sites, and with further nearest neighours coupling. iii)  The bath degrees of freedom are traced out exactly in the DVR basis, and a closed form expression for the averaged current  in the incoherent tunneling regime is obtained. Subsequently  our results are applied to investigate  current rectification and its {\em optimisation} in ac-driven    ratchets.

We  start considering the total Hamiltonian $H(t)=H_{\rm R}+H_{\rm ext}(t)+H_{\rm B}$, where  $H_{\rm R}=p^2/2m + V_{\rm R}(x)$ is the Hamiltonian for a particle of mass $m$ moving in the asymmetric periodic ratchet potential $V_{\rm R}(x+L)=V_{\rm R}( x)$, cf. Fig. 1. Due to Bloch theorem, the spectrum of the isolated system follows from the Schr\"odinger equation $ H_{\rm R} |\Psi_{n,k}\rangle ={\cal E}_n(k) |\Psi_{n,k}\rangle $,
with  $n$ denoting the band index, and with $k$ being the wave-vector. Time-reversal symmetry plus periodicity yield
 ${\cal E}_n(k)=E_n +\sum_{m=1}^\infty (\Delta_n^{(m)}/2)\cos(mkL)$.
 The action of the unbiased force is captured by $ H_{\rm ext}(t)= xF(t)$,  with $F(t)=F\cos\Omega t$, while  $H_{\rm B}$ is the standard Hamiltonian
for a harmonic oscillator bath with bilinear coupling in the bath and system coordinates \cite{Weiss99}.   The character of the bosonic bath is then fully described by the spectral function $J(\omega)$. Such a quantity is obtained from the equilibrium noise properties  of the real system to be described. We assume the Ohmic form, relevant e.g. in quasi one-dimensional Josephson junction arrays, $J(\omega)=\eta\omega/(1+(\omega/\omega_{\rm D})^2)$,  corresponding to a white  noise spectrum in the classical limit, with $\eta$ being the friction coefficent, and with $\omega_{\rm D}$   a bath cut-off frequency.

We introduce 
 Wannier states localized in cell $j$ as $|n,j\rangle=(\sqrt{L/2\pi})\int_{-\pi/L}^{\pi/L}dk e^{-ikjL}|\Psi_{n,k}\rangle$.
Due to  lack of inversion symmetry, it  holds  
$\langle x|j,n\rangle:=\Psi_{j,n}(x)\neq \Psi_{-j,n}(-x)$ for some $x$.
In the following, we  restrict our attention to the lowest $M$ bands of the potential.
 We focus on Hamiltonians whose spectrum is well approximated by tight-binding Hamiltonians with
 only nearest-neighbors coupling (miniband-approximation),  i.e., ${\cal E}_n(k)=E_n+(\Delta_n/2)\cos kL$.

 Then, the ratchet Hamiltonian is fully characterized by the $2M$ parameters, $\{\Delta_n,E_n\}$, $n=1,..,M$, determining the band widths, and the band centers, respectively.   It has the form in the Wannier representation
\begin{eqnarray}
 H_{\rm R}&=&\sum_{j=-\infty}^{\infty}\sum_{n=1}^M\frac{\Delta_n}{4}(|j,n\rangle\langle j+1,n|+ |j+1,n\rangle\langle j,n|)\nonumber\\
&+&\sum_{j=-\infty}^{\infty}\sum_{n=1}^M E_n|j,n\rangle\langle j,n|\;.
\label{HWannier}
\end{eqnarray}
Correspondingly, the discrete position operator reads
\begin{equation}
   x=
 \sum_{j=-\infty}^{\infty}\sum_{n,m=1}^M
x_j(n,m)|j,n\rangle\langle j,m|   \;,
\label{ext}\end{equation}
where, due to the localized character of the Wannier states, intercell matrix elements were neglected. Here,
  $x_j(n,m)= \langle j,m| x|j,n\rangle :=
jL\delta_{nm} +\xi_{nm}$, where the matrix  $\xi_{nm}=\xi_{mn}$ carries information on the {\em shape} of the ratchet potential $V_{\rm R}(x)$ within a cell.
 {Only} within the commonly performed lowest band truncation, $M=1$,   the position operator $  x$  is diagonal in the Wannier basis.

We  wish to evaluate the  average particle's velocity
\begin{equation}
 v=\langle v_{\rm as}(t)\rangle_\Omega:=\langle\lim_{t\to\infty}
{\rm Tr}\{ x\dot\rho(t)\}\rangle_\Omega\;.
\end{equation}
  Here $\rho(t)={\rm Tr}_{\rm Bath}W(t)$ is the
 reduced density matrix (RDM) of the system. It is obtained by performing the trace over the bath modes in  the
 density matrix $W(t)$   of the system-plus-reservoir. Finally,
$\langle ...\rangle_\Omega$ denotes the time average over a
driving period.
 The evaluation of the velocity $v$ is quite
 intricate: As clearly seen from (\ref{ext}), the coupling
 to the thermal bath  and to the external field modifies
 the quantum coherent Bloch picture in (\ref{HWannier}).
 Being the position operator  {\em not diagonal}
 in the Wannier basis,
 interband transitions become possible due  both to   the coupling
 to the thermal bath  and to the external field.

 For a harmonic bath  bilinearly
 coupled to the system, the trace over the bath degrees of freedom
 can be done exactly by use of real-time
 path-integral techniques.
 This requires, however, to express the full system-plus-bath Hamiltonian
 $H(t)$  in the basis of the eigenstates of the position operator,
 the so termed  discrete variable representation (DVR) \cite{DVR,Thorwart}.
  The contributions $H_{\rm ext}(t)$ and $H_{\rm B}$
 to $H(t)$ are { already diagonal in this basis}, but not 
  the isolated ratchet Hamiltonian. Diagonalization of the matrix $\xi_{nm}$  yields  the orthogonal transformation $U$ from the Wannier to the DVR representation, $|j,n\rangle =\sum_\mu U_{n,\mu}|j,\mu\rangle   $,
 as well as   the position operator eigenvalues and
  eigenvectors  $x|j,\mu\rangle =x_{j,\mu}|j,\mu\rangle  =(jL + x_\mu )|j,\mu\rangle  $. Here $\mu$ labels the DVR states {\em within } a cell.  Thus
 the ratchet Hamiltonian reads in the DVR basis
\begin{eqnarray}
\lefteqn{H_{\rm R} = \sum_{j=-\infty}^{\infty}\left(\sum_{\mu=1}^{M}\epsilon_\mu
|j,\mu \rangle\langle j,\mu|+
\sum_{\mu\neq\nu=1}^M \Delta_{\mu,\nu}^{\rm intra}|j, \mu \rangle\langle j,\nu|\right)}\nonumber\\
&&\!\!\!\!\!\!+\sum_{j=-\infty}^{\infty}\sum_{\mu,\nu=1}^M \Delta_{\mu,\nu}^{\rm inter}(|j,\mu \rangle\langle j+1,\nu|+|j+1,\mu\rangle\langle j,\nu|)\;,
\nonumber
\end{eqnarray}
where the on-site energies read $\epsilon_\mu\equiv \sum_n E_n U_{n\mu}U_{n\mu}$, while
$\Delta_{\mu,\nu}^{\rm intra}\equiv\sum_n E_n U_{n\mu}U_{n\nu}$ are related to the {\em intra}cell vibrational motion. Finally,  $\Delta_{\mu,\nu}^{\rm inter}\equiv\sum_n (\Delta_n/4) U_{n\mu}U_{n\nu}$ accounts for intercell tunneling.  Tunneling transitions to any DVR state in the neighboring cell are allowed. Additionally, because $H_{\rm ext}(t)$
 is diagonal in the DVR basis, the  Hamiltonian $H(t)$ has {time-dependent} on-site
energies $F(t)x_{j,\mu}$ which add to the intrinsic ones $\epsilon_\mu$.
We have now all the ingredients to evaluate the asymptotic  tunneling current. Its expression reads 
\begin{equation}
v_{\rm as}(t)=\lim_{t\to\infty}\sum_{j=-\infty}^{\infty}\sum_{\mu=1}^M x_{j,\mu}\dot P_{j,\mu}(t)\;,
\end{equation}
with  the $P_{j,\mu}(t)$ being the diagonal elements of the reduced density matrix in the DVR basis. They represent the probability of finding the particle in the state $|j,\mu\rangle$ at time $t$. In the following, we restrict our attention to the incoherent tunneling regime, reached at high enough temperatures and/or friction \cite{Weiss99}.
 Moreover, we consider  moderate-to-high frequencies $\Omega$, such that $\omega_0 \ge \Omega>\Gamma_{\nu,\mu}^{j,j'}$, with
$\Gamma_{\nu,\mu}^{j,j'}$ being the field-averaged transition rate from
DVR site $|j,\nu\rangle $ to $|j',\mu\rangle $.
 The upper bound, with $\hbar\omega_0=E_2-E_1 $,  is   needed because of
 consistency with the truncation of the original problem
 to the first $M$ bands.  The lower bound enables to approximate
 the evolution of the
 averaged occupation probabilities $\overline P_{j,\mu}(t):=
\langle P_{j,\mu}(t)\rangle_\Omega$ by the   coupled  equations
\begin{eqnarray}
\lefteqn{\frac{d\overline P_{j,\mu}(t)}{dt}=-\sum_{\nu\neq\mu}[\Gamma_{\mu,\nu}^{j,j}\overline P_{j,\mu}(t)
- \Gamma\ _{\nu ,\mu}^{j,j}\overline P_{j,\nu}(t)]}
\nonumber\\&-&
\sum_{\nu}[\Gamma_{\mu,\nu}^{j,j-1}+\Gamma_{\mu,\nu}^{j,j+1}]\overline P_{j,\mu}(t)
\nonumber\\
&+&
\sum_{\nu}[\Gamma_{\nu,\mu}^{j-1,j}\overline P_{j-1,\nu}(t)+\Gamma_{\nu,\mu}^{j+1,j}
\overline P_{j+1,\nu}(t)]\;.
\label{coupled}
\end{eqnarray}
Within an incoherent
tunneling description it is appropriate to
evaluate such rates  as an expansion in terms of the coupling
matrix elements $\Delta_{\mu,\nu}^{j,j'}$, with $\Delta_{\mu,\nu}^{j,j}=\Delta_{\mu,\nu}^{\rm intra}$ and $\Delta_{\mu,\nu}^{j,j\pm 1}=\Delta_{\mu,\nu}^{\rm inter}$.
 To lowest order we find
\begin{eqnarray}
\lefteqn{\Gamma_{\mu,\nu}^{j,j'}=
\left(\Delta_{\mu,\nu}^{j,j'}/\hbar\right)^2\int_{-\infty}^\infty
d\tau e^{-Q_{\mu,\nu}^{j,j'}(\tau)}} \nonumber\\ &\times&
J_0\left(\frac{2F(x_{j,\mu}-x_{j',\nu})}{\hbar\Omega}\sin\left(
\frac{\Omega
\tau}{2}\right)\right)e^{i(\epsilon_{\mu}-\epsilon_\nu)
\tau/\hbar} , \label{rates}
\end{eqnarray}
where the external field parameters enter the argument of the  zero-order Bessel function $J_0$. This yields a strong reduction of the rate (\ref{rates}), when $\epsilon_\mu-\epsilon_\nu=n\hbar\Omega$ and the ratio $F(x_{j,\mu}-x_{j',\nu})/\hbar\Omega$ hits a
zero of the Bessel function $J_n$. In the absence of dissipation,
 similar conditions would yield a dynamical localization of the
 particle \cite{Physrep,Kenkre}.
 However, 
 finite temperatures of the bath hinder this effect.
The
environmental influence is determined by the bath correlation function
$Q_{\mu,\nu}^{j,j'}(\tau) =\frac{(x_{j,\mu}-x_{j',\nu})^2}{2\pi\hbar}
\int_0^\infty d \omega \frac{J(\omega)}{\omega^2}[\coth(\hbar\beta\omega/2)(1-\cos\omega\tau) +i\sin\omega\tau]$ \cite{Weiss99}. Note that the dissipation strength depends on the { square distance} between the   DVR states involved.
It is
{\em position-dependent} through the  dimensionless coupling parameter
\begin{equation}
\alpha_{\mu,\nu}^{j,j'}=\frac{\eta}{2\pi\hbar}(x_{j,\mu}-x_{j',\nu})^2\;.
\label{alpha}
\end{equation}
For symmetric cosine shaped potentials a bath-induced localization at zero-temperatures has been predicted for dissipation parameters $\alpha = \eta L^2/(2\pi\hbar)>1$ \cite{Bray}. A similar tendence to  localization 
 occurs in our ratchet for large dissipation strengths $\alpha_{\mu,\nu}^{j,j\pm 1}>1$ and zero temperature. However, appropriately tuned ac-fields can
 help preventing localization.

The  stationary solutions of the coupled system of equations (\ref{coupled}) can  be found  upon Laplace transformation. Then, the
current assumes the compact form
\begin{equation}
v=L\sum_{\nu,\mu}p_\nu^\infty(\Gamma_{\nu,\mu}^{{\rm inter},f}-\Gamma_{\nu,\mu}^{{\rm inter},b})\;,
\label{current_final}
\end{equation}
where we introduced the forward and backward time-averaged rates
$\Gamma_{\mu,\nu}^{{\rm inter},f/b}:=\Gamma_{\mu,\nu}^{j,j\pm 1}$,
and the vibrational relaxation rates $\Gamma_{\mu,\nu}^{\rm intra}:=\Gamma_{\mu,\nu}^{j,j}$ ($\mu\neq\nu$).
 The asymptotic cell occupation probabilities $p_\mu^{\infty}:=\lim_{t\to\infty}\sum_j \overline P_{j,\mu}(t)$
  are expressed in terms of
  the averaged rates. For $M=1$, the current in (\ref{current_final}) is zero,
  as a consequence of time-reversal
 symmetry. Thus single-band quantum ratchets support no current, as recently 
reported in \cite{Majer01}. They  define
 a new class of  supersymmetric \cite{Reimann01} potentials in the quantum regime. 
At least two bands should contribute to the
 dynamics in order to have a non-vanishing current. For   $M=3$
 contributing bands (or  DVR states) we find
\begin{eqnarray}
p_{1}^{\infty}&=&\frac{\Gamma_2^{}\Gamma_3^{}-\Gamma_{2,3}^{}\Gamma_{3,2}^{}}{
\sum_{\nu=1}^3\sum_{\gamma>\nu=1}^3
(\Gamma_{\nu}^{}\Gamma_{\gamma}^{}-\Gamma_{\nu,\gamma}\Gamma_{\gamma,\nu})}\;,
\label{cyclic}
\end{eqnarray}
where
$\Gamma_\nu=\sum_{\mu\neq \nu}\Gamma_{\nu, \mu}$, and
$\Gamma_{\nu ,\mu}:=\Gamma_{\nu,\mu}^{{\rm inter},f}+\Gamma_{\nu,\mu}^{{\rm inter},b}+\Gamma_{\nu,\mu}^{\rm intra}.$
Likewise, $p_{2}^\infty$ and $p_3^\infty$ can be obtained from
(\ref{cyclic}) upon cyclic index permutation.
In the {\em absence} of ac-driving the  rates (\ref{rates})
obey the detailed-balance relation
\begin{equation}
\Gamma_{\mu,\nu}^{j,j'}=\Gamma_{\nu,\mu}^{j',j}e^{-(\epsilon_\mu-\epsilon_\nu)/k_{\rm B}T}\;,
\label{detailed-balance}
\end{equation}
so that the total current in (\ref{current_final}) adds up to zero.
The ac-field invalidates (\ref{detailed-balance}), and 
a net  current is originated.

Up to this point, our results are general. We apply now our theory to make  predictions for 
vortex motion in quantum ratchets devices. These could be realized with
rectangular \cite{Majer01} or circular \cite{Falo02} arrays of different-sized Josephson junctions.
Upon applying a magnetic field perpendicular to the array, current vortices are originated, whose dynamics is homologous to that of a quantum Brownian particle in a periodic ratchet potential \cite{Lobb83,Orlando91}. The particle's mass is $m=\Phi_0^2C/2a^2$, where $\Phi_0$ is the flux quantum, $C$ is the average capacitance of the junctions, and $a$ is the average junction distance  in the 
direction of motion.
The magnitude of the potential felt by the vortices is proportional to the Josephson energy $E_J$.
 The vortices can be put into motion upon injecting a current 
 $I=I_{\rm AC}\cos(\Omega t)$ into the array, which exerts a Lorentz-like force
on such particles. 
 At low vortex densities, the voltage drop $V$ across the array   is related to the vortex velocity by the second Josephson relation
\begin{equation}
V=sv\Phi_0/L\;,
\end{equation}
where $s$ is the one-dimensional vortex density.
%
Figures 2 and 3 illustrate our predictions for current rectification for the driven ratchet potential shown in Fig. 1.  
It supports only three bands below the barrier, being characterized by  $\Delta_1=-4.43\times 10^{-3}\hbar\omega_0$, $\Delta_2=6.68\times 10^{-2}\hbar\omega_0$, and $\Delta_3=-3.10\times 10^{-1}\hbar\omega_0$. The center of the bands read, $E_1=5.87 \times 10^{-1}\hbar\omega_0$, $E_2= 1.59 \hbar\omega_0$, and
$E_3=2.45 \hbar\omega_0$. This is the potential reported in
 \cite{Majer01} at a temperature of $T=50mK$, corresponding to a charging energy $E_J=429\mu$eV=$8\hbar\omega_0$.
We choose an average capacitance $C=2$fF, an average distance
 $a=L/3=2\mu m$, and frustation $s=0.277$.  The
 average junction resistance   $R $ is taken as a free parameter. It determines   the  friction parameter  $\eta\approx \frac{\Phi_0^2}{2R a^2}$, and the bath cut-off $\omega_{\rm D}\approx (R C)^{-1}$ \cite{Orlando91}.
In Figs. 2 and 3 we focus on two different values of the
dimensionless friction parameter $\alpha$, cf. below (\ref{alpha}), i.e.,
 $\alpha =0.5$ and $\alpha=2.0$ which
describe  a moderate and a strong damping regime,  respectively.
%
 These fully specifies the ratchet potential and the bath parameters. 
 Finally,  the injected  ac-current 
 induces  a force $F(t)=I_{\rm AC}\cos(\Omega t)\Phi_0/N_c a$,
where $N_c=304$ is the  chosen number of junction columns of the array.
\begin{figure}
\epsfig{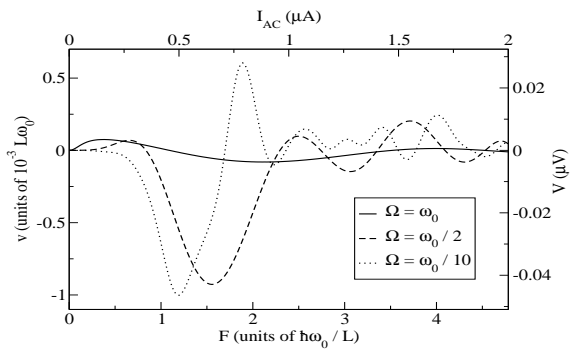}
\caption{Output velocity (or dc-voltage for the vortex ratchet
case) vs.  amplitude of the driving ac-signal (ac-current for
vortex ratchet) for different driving frequencies $\Omega$.  The
dimensionless friction parameter $\alpha=2\pi\eta L^2/\hbar$ is choosen to
be $\alpha=0.5$, describing moderate damping. Negative or positive
ratchet currents are obtained upon appropriate tuning of the
driving field parameters. }
\end{figure}
\begin{figure}
\epsfig{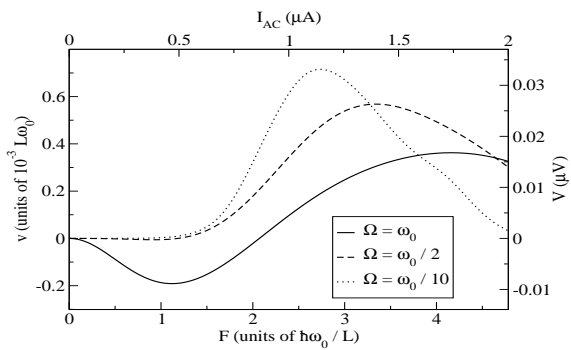}
\caption{Same as in Fig. 2 for  larger friction,  $\alpha=2.0$.
The main resonances occur at larger driving amplitudes.}
\end{figure}
Figs. 2 and 3 reveal how the  interplay among driving-induced
localization near the zeros of the Bessel function in (\ref{rates}),
driving-assisted tunneling when the driving frequency matches
resonances  of the dissipative ratchet system, and dissipation results in a
nontrivial {\em non-monotonic} dependence of the current on the applied
ac-amplitude and frequency. Likewise a non-monotonic behavior occurs varying
 damping. For example, different current directions are observed for small ac-amplitudes in the curves with  driving $\Omega=\omega_0$ of Fig. 2 and Fig. 3. This inversion
also occurs in combination with  a change in the bath cut-off frequency $\omega_{\rm D}$. As damping is further increased 
the main resonance peaks get shifted to even larger values of the driving amplitude, and the ratchet signal is reduced.

In conclusion we investigated the  ratchet mechanism in periodic quantum  structures with  {\em few} bands below the barrier, and driven by {\em non}-adiabatic
time-dependent fields. 
 A set of coupled master equations for the reduced density matrix elements in the basis which diagonalizes the position operator are derived.
 The validity of our method  is restricted to the deep quantum regime, since thermally activated processes over the barrier are disregarded. Our approach is  complementary to the semiclassical analysis 
 of \cite{Reimann97}, where many bands below the barrier are required. 
We discussed current rectification in Josephson ratchets.
Current reversals  result from the interplay of driving-induced
dynamical-localization \cite{Physrep} versus dissipation-assisted diffusion.
Optimal current rectification is obtained for moderate damping, and driving
frequencies of the order of the typical relaxation time of the environment.

\centerline{***}
We thank M. Thorwart, P. H\"anggi and J.E. Mooij for useful discussions.
Support by FOM is acknowledged.

%
%
%
\end{document}